\documentclass[runningheads]{llncs}
\usepackage[T1]{fontenc}
\usepackage{graphicx}
\begin{document}
\title{Comparative Analysis of STEM and non-STEM Teachers’ Needs
for Integrating AI into Educational Environments}
\author{Bahare Riahi\inst{1} \,
Veronica Cateté\inst{2}}
\authorrunning{B. Riahi and V. Cateté} % Shortened names for running head
\institute{North Carolina State University, Raleigh, NC 27606, USA \\
\email{briahi@ncsu.edu} 
\email{vmcatete@ncsu.edu}}
\maketitle              % typeset the header of the contribution
\begin{abstract}
There is an increasing imperative to integrate programming platforms within AI frameworks to enhance educational tasks for both teachers and students. However, commonly used platforms such as Code.org, Scratch, and Snap fall short of providing the desired AI features and lack adaptability for interdisciplinary applications. This study explores how educational platforms can be improved by incorporating AI and analytics features to create more effective learning environments across various subjects and domains.

We interviewed 8 K-12 teachers and asked their practices and needs while using any block-based programming (BBP) platform in their classes. We asked for their approaches in assessment, course development and expansion of resources, and student monitoring in their classes. Thematic analysis of the interview transcripts revealed both commonalities and differences in the AI tools needed between the STEM and non-STEM groups. Our results indicated advanced AI features that could promote BBP platforms. Both groups stressed the need for integrity and plagiarism checks, AI adaptability, customized rubrics, and detailed feedback in assessments. Non-STEM teachers also emphasized the importance of creative assignments and qualitative assessments. Regarding resource development, both AI tools desired for updating curricula, tutoring libraries, and generative AI features. Non-STEM teachers were particularly interested in supporting creative endeavors, such as art simulations. For student monitoring, both groups prioritized desktop control, daily tracking, behavior monitoring, and distraction prevention tools.

Our findings identify specific AI-enhanced features needed by K-12 teachers across various disciplines and lay the foundation for creating more efficient, personalized, and engaging educational experiences.

\end{abstract}
\section{Introduction}
It is no longer assumed that managing classroom and teaching activities in computational teaching and programming instruction follows a standardized approach \cite{koulouri2014teaching}, with fixed strategies and schedules for instructors. Various factors influence teachers' strategies, including course type, differences in students' proficiency levels \cite{figueiredo2021strategies}, teachers' conceptual and educational backgrounds, experience \cite{leyzberg2017teaching,huang2022comparison}, and methodologies \cite{compan2019effects}.

Given the complexity and variety of teachers' activities in classroom management and enhancing students' learning, supplementary tools are essential \cite{rossling2008enhancing,mahtani2022online}. These tools support tasks such as instruction and assessments, including setting rubrics, projects, and quizzes \cite{milliken2021exploring}. Prior research has explored teachers' general needs for programming learning systems \cite{limke2024survey} and developed tools for block-based programming (BBP) environments, such as assignment hubs, grading tools, and planning interfaces \cite{harvey2013snap,milliken2021exploring}. However, these needs vary significantly across instructional fields, especially in non-computing or non-STEM courses.

We aim to identify teachers' needs and preferences regarding the integration of AI features in educational platforms (EP), with a focus on three key areas: assessment; course development and resource expansion; and monitoring and supporting students. Additionally, we address the research gap concerning non-STEM teachers' needs in using AI in block-based programming (BBP). Recognizing this gap we explored how AI could be integrated into educational platforms to enhance its accessibility and effectiveness for a broader range of educational needs, spanning both STEM (Science, Technology, Engineering, and Mathematics) and non-STEM (Arts, Social Studies, and Humanities) disciplines \cite{uddin2021research}.
Thus, ensuring the adaptability and accessibility of these platforms is of fundamental importance. Additionally, identifying customized AI analytics features to meet the diverse needs of users is essential for creating more effective learning environments

To achieve these objectives, we conducted semi-structured interviews with eight K-12 teachers—four from STEM and four from non-STEM disciplines—to understand their approaches to assessment, course development and resource expansion, and student monitoring and support. We explored their methods, strategies, current use of AI tools, and the features they would like to see added. Thematic analysis was used to analyze their responses for each focus area.
For course development and expansion, teachers highlighted the need for built-in code tracing, dynamic document updates, and customized course materials tailored to individual student needs and levels. Additionally, they emphasized the importance of monitoring features such as customizable tools for individualized growth tracking, help notifications, pop-up tips, reminders, and motivational tools.

\section{Related Works}
With widespread advancements in technology and computer science, there is a growing need to educate students in computer science and computational thinking from an early age \cite{tekdal2021trends}. With the emphasis on developing computational thinking (CT) and creativity in K-12 education increasing \cite{lee2022computer}, and considering benefits such as improved future college performance \cite{burgiel2020association}, more schools are incorporating computer science as a core subject or integrating it into their curricula. Computational thinking (CT), which was introduced by Wing \cite{wing2006computational}, is the ability to solve problems, create solutions through abstraction and algorithmic thinking, and comprehend human behavior by utilizing the core principles of computer science \cite{chen2017assessing}.

There has been growing interest in learning CT for all students regardless of their discipline \cite{hsu2018learn}, as it can be integrated into both STEM (Science, Technology, Engineering, and Mathematics) and non-STEM (Human Societies, Language, Communication and Culture, and History and Art) \cite{liao2022exploring} disciplines. In K-12 classes, students use block-based programming (BBP) using various tools, such as Code.org \cite{kaleliouglu2015new} to develop their CT skills. 
The adoption of these tools reflects a wider movement in education, in which supplementary technologies and AI are used to increase classroom efficiency and productivity \cite{fitria2021artificial}. These innovations not only simplify administrative duties but also enhance student engagement and allow for personalized learning experiences. Considering the ongoing challenges in performing teaching tasks, there is a significant demand for such tools to support teachers with curriculum-focused activities, instruction, student progress monitoring, and the creation of thorough assessments. 

AI tools play a pivotal role in supporting K-12 computer science educators by streamlining assessment processes \cite{milliken2021exploring} and providing substantial assistance to both teachers and students \cite{milliken2021planit}. As there is a relatively small body of literature focused on integrating AI tools into block-based programming (BBP) in both STEM and non-STEM classes, we argue that the complexities of diverse assessment modules and distinct pedagogical approaches can pose significant challenges for teachers \cite{milliken2021exploring}. Moreover, AI can customize learning experiences by adjusting content and progression to meet students' specific needs. By leveraging AI in these ways, educators can enhance their ability to provide effective performance and supportive feedback to their students \cite{lin2021engaging}.

\begin{table}
\caption{Details of the participants, teaching Grade level and their fields in both groups
of STEM and non-STEM}\label{tab1}
\begin{tabular}{|p{4cm}|p{2cm}|p{4cm}|p{2cm}|}
\hline
non-STEM Teachers (NST) & Grade Level & STEM Teachers (ST) & Grade Level\\

\hline
Art-3D &  {9-12} & CS principles/ math & 9-12 \\

English (ESL) &  {11} & Computer Science & 11-12 \\
Music & 9-12 & Math & 10  \\
Dance & 6-8 & Math/Science/IT & 10 \\

\hline
\end{tabular}
\end{table}

\subsection{Integration block based programming in non-stem classes}
BBP has gained significant attention in various educational settings, including non-STEM classes such as social science, humanities, dance, English for Speakers of Other Languages (ESOL), and art \cite{sullivan2017dancing,guzdial2024creating}. Incorporating BBP in non-STEM classes can be beneficial for students to enhance their computational thinking abilities and develop skills that are essential in the digital age. Scholars showed that Integrating programming into art and music education can significantly enhance the learning experience, making it more appealing \cite{grassl2024girls}.

Using BBP programming in interdisciplinary and non-STEM areas like Arts \& Crafts and Music, using tools like Scratch \cite{marji2014learn}, reduces barriers for teachers and students, enhances engagement, increases the creativity of students and practical applications of programming across different subjects—making learning more exciting and accessible for everyone \cite{perin2023investigating}. Given the need for educational platforms to be accessible across both STEM and non-STEM subjects, and the importance of facilitating class management for teachers, it's crucial that these platforms include features that meet the diverse requirements of all user groups.

\subsubsection{Art, Music, Dance and English for Speakers of Other Languages (ESOL) Education}
Students use educational visual programming tools like Scratch \cite{begosso2020analysis} and Snap! to enhance their musical understanding and compositional skills in subjects like music and dance. Integrating block-based programming with dance not only improves students' computational skills and artistic abilities but also provides teachers with effective ways to assess students' progress. By using sound and music blocks, students can compose melodies, manipulate pitch, create rhythmic patterns with loops and conditional statements, and deepen their comprehension of musical timing and structures \cite{shamir2019paradigm}.

Additionally, through musical BBP, students can animate characters and synchronize their movements with music using motion and sound blocks, enabling dynamic dance choreography that responds to musical cues or user interactions \cite{bi2018real}. Scholars have emphasized that 3D programming can further enhance creativity in computational thinking \cite{repenning2014beyond}, and using Scratch to animate dance moves and synchronize music significantly improves students' understanding of computational thinking in the context of dance, music, and art \cite{shamir2019paradigm}.

BBP can be used to enhance foreign language learning and spelling skills \cite{holz2020design} among students. It has been shown that by using jigsaw-like placement of colored blocks, students can explore and understand grammar structures and vocabulary, get immediate feedback and correct their mistakes, and make different constructions of the sentence \cite{purgina2020wordbricks}. There is a growing need to support and equip \cite{yim2024artificial} teachers in teaching and managing their classes more effectively.
By applying various BBP tools in these areas, teachers encounter different needs specific to their subjects. To improve accessibility and keep these tools up-to-date, we aim to explore teachers' requirements for adding AI features, particularly in non-STEM areas.

\section{Research Questions}
We aim to answer our research questions in the three area of assessment, course development and student monitoring. 
Our first research question explored how to better address the assessment needs of STEM and non-STEM teachers across various domains.
\begin{itemize}
    \item[\textbf{RQ1}] What AI features do STEM and non-STEM teachers need and want in the educational tools for student assessments?
\end{itemize}
Another important consideration is how well the dashboard aligns with curricula, teaching approaches, and the needs of various subject areas. Therefore, the second research question is:
\begin{itemize}
    \item[\textbf{RQ2}] What AI features do STEM and non-STEM teachers need and want for educational tools to expand their resources?
\end{itemize}
Research has shown that, in computer science courses, teachers are actively involved in monitoring and supporting student and improve their progress \cite{dong2021using,dong2021you}.
Therefore, our study also explores how AI tools can effectively address the needs of teachers in monitoring students’ struggles and providing support:
\begin{itemize}
    \item[\textbf{RQ3}] What AI features do STEM and non-STEM teachers need and want in educational tools to monitor and support students?
\end{itemize}

\section{Methods}
\subsection{ Data Analysis}
In the initial phase of our research, we recruited eight K-12 teachers from schools in North and South Carolina in the United States. We conducted semi-structured interviews with both STEM and non-STEM teachers who used AI in their classes. The inclusion criterion for the interview participants was that they had previously implemented coding activities in their STEM or non-STEM classrooms. These activities ranged from short-hour-long modules to four-day lessons that followed a use-modify-create progression \cite{franklin2020analysis}, allowing students to complete open-ended coding assignments. 
We used a pre-survey to select teachers based on their grade level and experience by incorporating coding into their subject areas. We selected eight participants, which have the requirements, (four from each group). The interview questions categorized in to three sections: assessment, resource planing and student monitoring. We asked about the way they perform assessment and how they expand their resources, how they support students, and how they monitor their struggles in their classes. Moreover, we asked them how they have been using educational platforms, if any, to perform the activities mentioned in their classes and if they are integrated with AI features, which features is their favorite. Finally, we asked them about their expectations from AI to enhance the belated platforms and what features they would like to see added to meet their needs better.

Each interview lasted between 40 minutes and one hour, and participants were compensated with a \$40 gift card. The interviews were conducted via Zoom, a video conferencing software, between March to May 2024 to accommodate the teachers' schedules. According to institutional review board (IRB) regulations, our study qualified as exempt from IRB review. 
All interviews were recorded with the teachers’ consent, and participants provided explicit consent at the start of each interview for recording the sessions for data analysis purposes. The recorded videos were accessible only to the researcher.
Considering the complex and varied factors, such as differences in course types (e.g., computing or non-computing), teachers' backgrounds \cite{tagare2024factors}, experiences, and students' individual differences, we chose a qualitative method via interview data collection and applied thematic analysis \cite{alhojailan2012thematic,maguire2017doing} to our transcripts to uncover in-depth insights into teachers' needs for AI tools \cite{johri2014conducting}. 

The analysis process involved reviewing the recorded videos, transcribing the interviews, and thoroughly reading the transcripts to explore the data comprehensively. During this process, we took notes and generated initial codes, which were grouped into broader categories, referred to as themes, to address our research questions. Two reviewers independently coded the data and compared their findings. Any code identified by one reviewer but missed by the other was added to the final list. These finalized codes were then grouped into thematic categories.
We categorized themes based on teachers' perspectives in each section of the interview: \textbf{a)} perspectives on adding AI features to current grading and assessment tools, \textbf{b)} perspectives on adding AI features to enhance existing tools, and \textbf{c)} perspectives on adding AI features to improve monitoring of students’ progress. We compared our tags and identified some that are common between STEM and non-STEM teachers, along with some unique themes. We use 'ST' and 'NST' as abbreviations for STEM and non-STEM teachers, respectively, followed by a number to indicate specific individuals.

% \begin{figure*} [h!]
%     \centering
%     \includegraphics[width=2 \columnwidth]{Assessment.png}
%     \caption{Percentage of Stem and non-STEM teachers need AI features for Assessment}
%     \label{SnapClass}
% \end{figure*}
% \begin{figure*} [h!]
%     \centering
%     \includegraphics[width=2 \columnwidth]{Resources.png}
%     \caption{Percentage of STEM and non-STEM teachers need AI features for course development and expanding resources}
%     \label{SnapClass}
% \end{figure*}
% \begin{figure*} [h!]
%     \centering
%     \includegraphics[width=2 \columnwidth]{Monitoring.png}
%     \caption{Percentage of STEM and non-STEM teachers need AI features for Students Monitoring }
%     \label{SnapClass}
% \end{figure*}

\section{Results}

% \subsection{Evaluation For Assessment}

% We found that STEM teachers often use formative assessments, such as quizzes in group projects and final tests. In contrast, non-STEM teachers favor more qualitative and non-rubric-based assessments, such as portfolios and project visualizations, with a focus on independent work and manual feedback. Both groups use checklists, rubrics, and online assessment systems and include projects in their final assessments.

%     \begin{figure*} [h!]
%     \centering
%     \includegraphics[width=1 \columnwidth]{Assess.png}
%     \caption{Percentage of Teachers (STEM \& non-STEM) Needing Assessment Features}
%     \end{figure*}
    
\subsection{Evaluation For Assessment: AI Features Needed By Teachers}
We found that STEM teachers often use formative assessments, such as quizzes in group projects and final tests. In contrast, non-STEM teachers favor more qualitative and non-rubric-based assessments, such as portfolios and project visualizations, with a focus on independent work and manual feedback. Both groups use checklists, rubrics, and online assessment systems and include projects in their final assessments.

    \begin{figure*} [h!]
    \centering
    \includegraphics[width=1 \columnwidth]{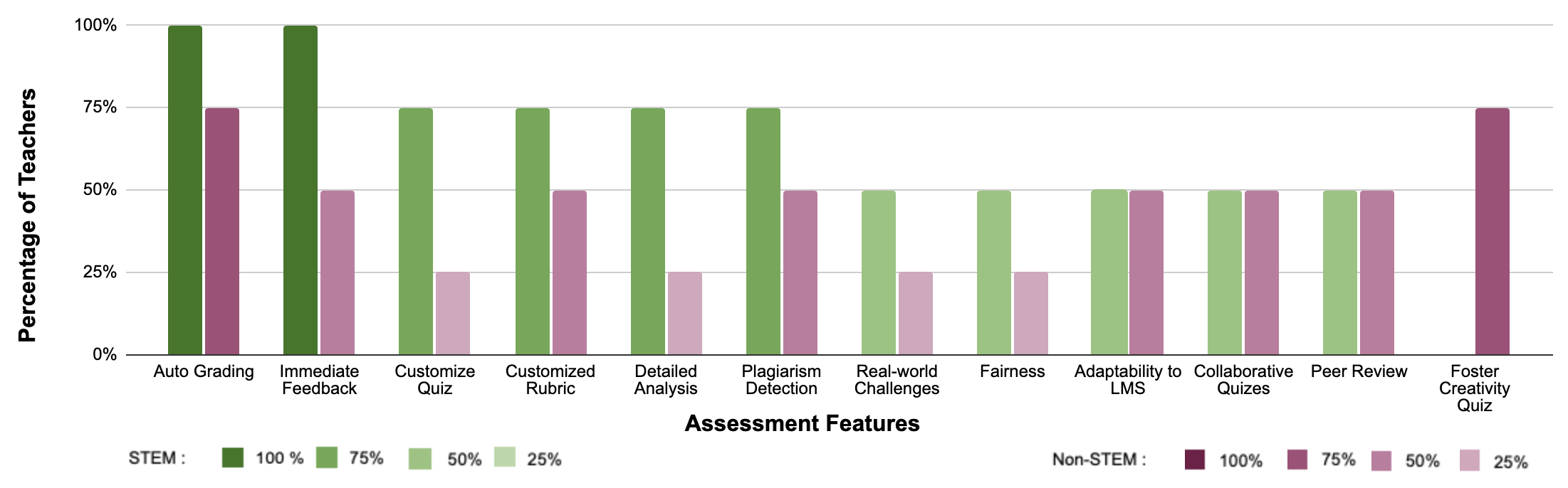}
    \caption{Percentage of Teachers (STEM \& non-STEM) Needing Assessment Features}
    \end{figure*}
    
After analyzing the themes based on the transcript we categorize them based on our research questions. Exploring RQ1 we found the following features that teachers in both groups need and want for student assessment. 

% Customization Of Rubrics and Individualized Assessment, Enhanced Feedback and Supportive Learning Resources, Originality and Authenticity, Integration and Adaptability to LMS st3,nst2

\subsubsection{Customization Of Rubrics and Individualized Assessment}
While current tools can create and customize questions based on curricula, enabling teachers to manipulate rubrics is essential for fostering student growth. Additionally, teachers require more individualized capabilities to address students' specific learning levels. For example ST1 "STEM teacher number1" mentioned \textit{I wish I could add some features to my current platform and change the rubrics whenever needed and based on everyone needs. It’s not just about getting the right answer—it’s about seeing their progress, capability and their thought and giving them feedback that helps them improve step by step}. Such an ideal platform should offer deeper customization options and effectively measure students' progress. In the feedback section, formative assessment—a key strategy widely used by STEM teachers—was highlighted as an essential feature. 

\subsubsection{Enhanced Feedback and Supportive Learning Resources}
Although existing platforms can trace code and grade based on rubrics or answer correctness, more detailed and partial grading is needed for line-by-line code assessment. As mentioned by ST3, teachers need to provide more detailed feedback, similar to tools like Quizizz, which offer immediate feedback and hints for each individual answer. They prefer students to understand and see their feedback even for a single line of code. More importantly, NST1 noted that after viewing their feedback and grades, students should have access to relevant and recommended tutorials, including documents, instructions, videos, and other resources to help them correct their mistakes. This approach prevents students from being overwhelmed by having to study all materials, thus reducing their workload. This type of structured feedback and support is especially crucial in visual arts classes, which often involve independent work.

\subsubsection{Originality and Authenticity}

Non-STEM teachers emphasize AI generative integration into EP that preserves the originality and authenticity of students' work and incorporates simulation capabilities for assessments in fields like dance and art. They seek functionalities that can evaluate creative assignments and create real-world challenges by converting real-world projects into digital grading systems. Given the diverse nature of their courses—such as art, dance these teachers prioritize features that support the authenticity of students' work. For example, NST1, who teaches art, highlighted that \textit{“originality and authenticity are the main requirements in art-related courses.”}

Additionally, in courses like dance, art and design, simulation is a crucial component of assessment. There is a clear need for tools that support simulation and enable evaluation based on simulations and templates. For instance, NST4, who specializes in dance, mentioned, \textit{“I create a simulation of dancing with Snap and also use Scratch to animate the dance movement.”} NST4 also pointed out the need for online assessment tools that can evaluate simulations and convert real-world work into software-based grading.
Additionally, three STEM teachers expressed concerns about the authenticity of their students' code and assignments. For example, ST2 stated that \textit{" although I encourage students for peer learning in my classes, some students using online resources and do copy-pasting. I need a feature that not allowed them to do so or show me parts of codes that weren't written by themselves."}

\subsubsection{Integration and Adaptability to LMS (Learning Management System) and Simplified Access Control}
Switching between platforms and transferring information such as feedback, grades, and student progress data requires significant time and effort. The Participants expressed a need for better integration and adaptability among the tools they use. For instance, ST1 highlighted their use of the generative AI tool Copilot, built into the Microsoft Edge browser, to create rubrics and instructional materials. They emphasized the necessity for seamless integration of AI-generated content (e.g., from ChatGPT and Copilot) into educational platforms like as Google Classroom.
Additionally, they prefer more comprehensive tools that encompass all necessary features to streamline these tasks. Grading, one of the most burdensome and time-consuming tasks for teachers, can be greatly facilitated by such integrated tools \cite{milliken2021redesigning}.

Moreover, one important feature, according to ST1, is SSO (Single Sign-On) to simplify access control and enhance security, it also simplifies managing the dashboard for both teachers and students. Therefore, this feature could enhance the usability of the platforms. ST2 mentioned, \newline \indent \textit{“Once students use the platforms such as scratch, if these platforms were compatible with SSO support features, students can log in to their Google Classroom account, and from there, they can access other third party platforms, like snap, scratch whatever it is. It has a lot of benefits, besides easy access, all students' data will be managed under a single authentication system”.}
\newline The mentioned STEM teachers have reported that their current platforms (such as syncing HMH with Canva or Admenton with each other) do not integrate seamlessly with school platforms and Google Docs for reporting final grades and feedback. On the other hand, NST1 and NST3 declared accessing different platforms for learning and visualizing progress is essential. Moreover, they articulated the need for features that simplify access to various resources, such as links to music practice exercises, guidance on correcting techniques, and step-by-step guides, which would be highly beneficial. Therefore, adaptability and flexibility are crucial AI generative abilities for those platforms to use in classroom management.

\subsection{Evaluation For Course Development \& Expanding Resources: AI Features needed by teachers}
Exploring RQ2a, we found that teachers highlighted the need for AI tools to support the following:
STEM teachers have identified the need for AI features integration that assist with creating documents for each topic and generating curricula. 

Both STEM and non-STEM teachers mentioned that AI features integration would be useful for developing course materials, including lesson plans, course descriptions and image generation. These features include but are not limited to automated content generation, personalized learning pathways, predictive analytics, resource optimization, and enhanced collaboration.

These features can enhance the platforms by analyzing students' performance and historical data to create personalized learning pathways and predict trends. This can lead to showing the course materials in an individualized way to students and changing the order based on their priorities based on their needs. This content adjustment and customized resource allocation ensure that course materials are effectively utilized and available as needed.
For example, ST1 and ST2 mentioned, \textit{“Sometimes students get mixed up about which course materials are the most important, especially when they're falling behind their classmates for one reason or another. They really need a bit of guidance to point them towards the materials they should focus on.”}

On the other hand, NST1, who teaches art, noted: \newline \indent \textit{“Since I teach art, simulating final artwork by uploading images and generating images with AI features would be enticing for me and my students. However, AI-generated images often lack originality and human creativity. While these tools enhance students' self-confidence and provide a basic idea, I hope AI can further inspire students to be more creative”.}

\begin{figure*} [h!]
    \centering
    \includegraphics[width=1 \columnwidth]{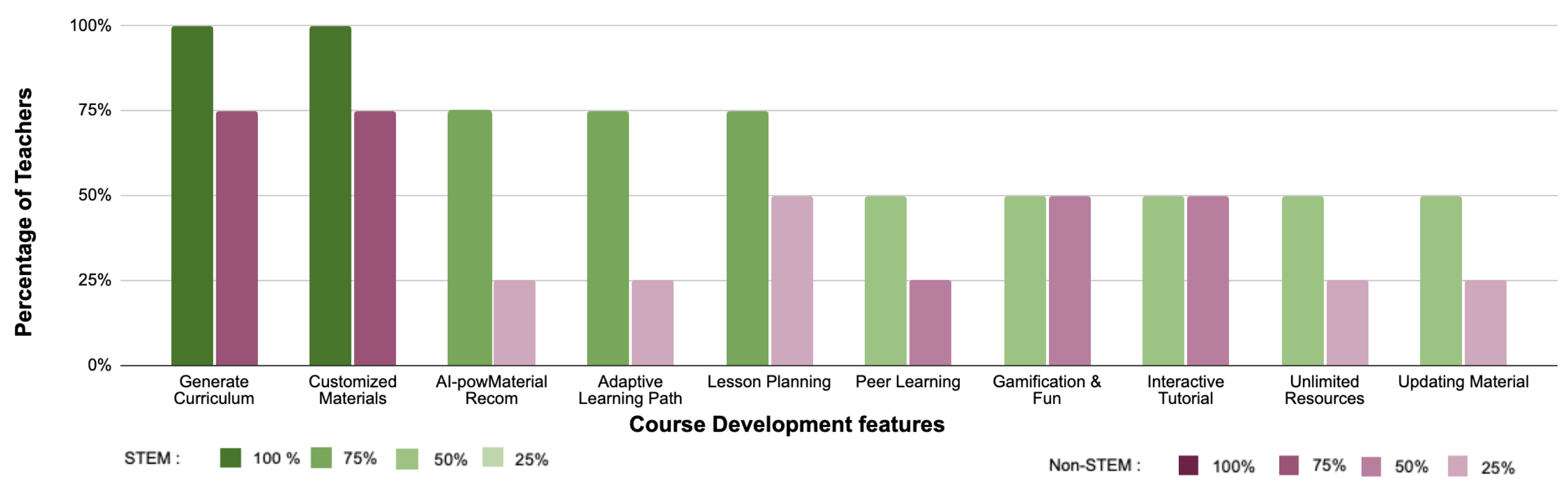}
    \caption{Percentage of Teachers (STEM \& non-STEM) Needing Course Development Features}
\end{figure*}

\subsubsection{Individualized (customized materials) and Accessible for \textbf{E}nglish 
\textbf{L}anguage \textbf{L}earners}
NST3 emphasized the importance of enhancing usability, suggesting that the platform would be more effective if it were smarter, with features like automated language level assessment, curriculum adjustments based on skills that need reinforcement, and optimized distribution of learning materials according to students' individual learning needs. NST3 addressed the challenges faced by non-English-speaking students using the Brisk Teaching platform and expressed a desire for the platform to tailor English instruction to individual levels, as well as facilitate the remote distribution of materials and quizzes for continuous learning.

\subsubsection{Peer Review and Collaborative Learning}
One factor both STEM and non-STEM teachers highlighted, though in different ways, is the importance of collaborative learning. 
Teachers highlighted the benefits of students collaborating on coding projects, such as discussing ideas, explaining their reasoning, and negotiating solutions. They noted that collaboration enhances problem-solving by fostering creative solutions and a deeper understanding of concepts. Additionally, students can learn effectively from their peers and teach each other strategies and solutions.

They wish AI had the following features: AI could analyze students' skills and performance and optimize peer matching to ensure teammates would have a balanced mix of abilities or could help each other effectively. Moreover, in terms of solving conflicts and disagreements, AI could provide assistance by offering resolutions. As ST1 mentioned, \textit{“Although CodeCombat is highly engaging for pair work, I wish AI features could intervene to resolve disputes among students, facilitate peer feedback, and help students be more creative”.}

\subsubsection{Real-World Connectivity}

Another feature appreciated by teacher ST4 is the tool's ability to connect with the real world and provide interactive experiences. For example, being able to test their codes with devices mounting sensors that interact with their code such as LEGO Mindstorms \cite{moraiti2022coding}.
As ST4 noted:
\textit{“Using resources like circuit boards that come with a hands-on component or project base is fantastic and engaging for students. I would prefer to use such resources because they make learning more interactive and enjoyable.”}

\subsubsection{Gamification in Educational Tools}

Another feature that most current tools incorporate is gamification. Gamification has demonstrated significant potential in improving learning outcomes in K-12 education and positively impacts student learning \cite{buckley2016gamification,dehghanzadeh2024using}. It also increases motivation and engagement by providing clear goals and a structured learning process \cite{choi2024exploring}. Participants NST1, NST4, ST2, and ST3 have experience using tools such as Kahoot, Quizizz, Spheros, Scratch, and Code.org to teach BBP. For instance, Code.org utilizes block-based coding and integrates gamification elements, enabling students to learn programming through game-based challenges, which enhances their motivation to learn coding \cite{choi2024exploring}.
Based on our participants' statements Integrate AI into interview would enhance usability by incorporating features such as adaptive learning challenges within games, rewarding students accordingly, fostering competitive yet collaborative learning and matching peers effectively on platforms.

\subsection{Evaluation of Student Monitoring}
When we analyzed teachers' responses about the educational platforms they use to monitoring students strugles, we found that they utilize platforms such as Edmentum, Quizizz and other school-related software. These tools provides real-time analytics, allowing teachers to monitor student performance on assessments.
Moreover, teachers discussed features that can track students' progress to closely evaluate and identify specific areas where a student may be struggling would be a great help.

Exploring RQ3, teachers from both groups expressed a desire for more control over their students' activities to prevent distractions. They are interested in built-in features that allow them to mirror students’ screens and control desktops, including the ability to block other tabs.
Teachers also wish for help notification suggestions, pop-up tips, deadline reminders, and features to increase student motivation by simulating progress alerts and notifying them of their classmates' study time.
\subsubsection{Accommodations for Individualized needs}
Teachers including ST1, ST3, NS2 have expressed their expectation that AI can be utilized to monitor and accommodate students with special needs by customizing tools for more effective student tracking. For example, NST2 mentioned, \textit{“A feature that can adjust expectations based on students' improvement and level, providing them with individualized plans, would be helpful”}.
ST4 explained, \textit{“I wish to have productivity tools that could identify where students are stuck and offer assistance, with pop-ups if they need help while doing their assignments”.}

  \begin{figure*} [h!]
    \centering
    \includegraphics[width=1 \columnwidth]{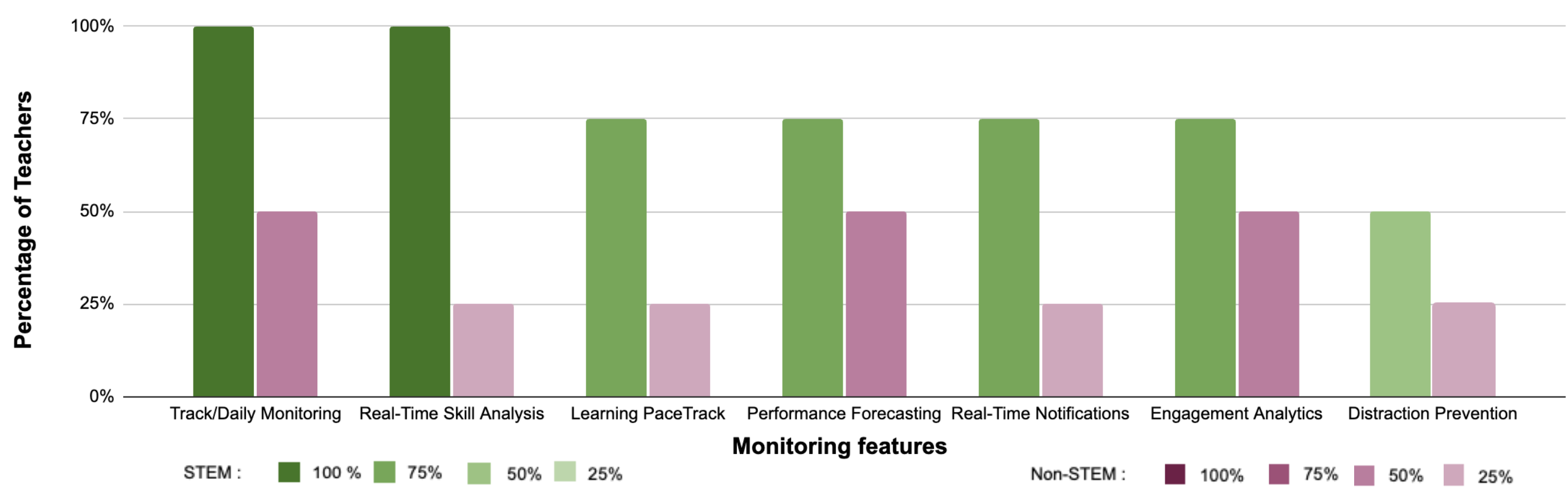}
    \caption{Percentage of Teachers (STEM \& non-STEM) Needing Monitoring Features}
    \end{figure*}

\section{Concerns regarding integrating AI into Educational Platforms}

\subsection{Privacy}
Here are some additional concerns teachers mentioned during their interviews that, while unrelated to the three main sections, are still noteworthy. One significant issue is privacy. Teachers expect AI platforms to include customizable security and privacy settings that comply with school policies. This would enable educators to safely integrate AI into classrooms and platforms while safeguarding student data and adhering to privacy regulations.
ST1 emphasized that AI tools must adhere to strict privacy requirements; failure to meet these standards could result in their usage being restricted or blocked. He mentioned, \textit{"the big challenge with using AI platforms in schools is nsuring ethical use and protecting student data. When Facebook acquired Oculus, we saw how a lack of adjustable security settings led to schools losing access due to data harvesting concerns. The same thing could happen with AI tools if they don’t meet strict privacy standards. We need AI companies to provide flexible security and privacy settings so we can safely integrate these tools into classrooms. Privacy is our top concern—just look at how schools restrict access to DALL-E because of these limitations"}.
Therefore, ensuring that these platforms adhere to strict privacy guidelines to protect individuals in all online platforms is essential \cite{behfar2024first}. It is not only important for fostering trust among educators but also for ensuring that the integration of AI technologies into educational settings is both sustainable and compliant. This focus on privacy and security will be vital in enabling broader acceptance and use of AI in classrooms without compromising student data.

\section{Discussion}
When comparing STEM and non-STEM teachers, it becomes clear that they integrate educational platforms differently into their teaching and have distinct expectations for AI integration into these platforms. STEM teachers typically follow structured curricula with specific content goals and are accustomed to using technology, often having already integrated advanced systems into their practices.

In contrast, non-STEM teachers, particularly those in the humanities and arts, often have more flexibility in their teaching plans. Their curricula may not be as rigidly defined by standards, allowing for creativity and variation. These teachers may not have prior experience using technology like computers and AI in their classrooms.
The challenge lies in addressing these knowledge gaps. Non-STEM teachers may require more practical examples and personalized support to grasp how AI can benefit their teaching practices. Due to technological familiarity, teacher expertise and resource utilization, non-stem teachers require more support in using technologies. Therefore,they may be starting from scratch when it comes to integrating AI technology.z

\section{Conclusion}
Through semi-structured interviews with 8 K-12 STEM and non-STEM teachers, we explored preferences to integrate AI tools into curricula and coding activities. We examined the intricacies of student assessment, course development and progress monitoring across different subject areas. We found valuable insight of what are the AI features that are needed by STEM and non-STEM teachers in the mentioned areas. 
Our findings indicate that both STEM and non-STEM teachers require significant improvement to current features. Many changes are feasible with recent advances in AI, such as developing analytical features for tracking students improvements and their status of progress, more detail feedback and customized course materials. However, some needs, such as control of student devices, are out of scope, and alternative means of student engagement such as gamification may be explored.

\end{document}